\documentclass{aa}

\newcommand{\unit}[1]{\; \mbox{#1}}

\usepackage{color}
\usepackage{amssymb}
\usepackage{graphicx}
\usepackage{aas_macros,natbib}

\definecolor{grey}{cmyk}{0.,0.,0.,0.5}

\begin{document}

\title{Spatial Origin of Galactic Cosmic Rays in Diffusion Models: 
II- Exotic Primary Cosmic Rays}
\author{D. Maurin\inst{1,2}
          \and R. Taillet\inst{1,3}}

          \authorrunning{Maurin \& Taillet}
          \titlerunning{Spatial Origin of Cosmic Rays in Diffusion Models}
          \institute{Laboratoire de Physique Th\'eorique {\sc lapth},
          Annecy--le--Vieux, 74941, France
          \and
	  Institut d'Astrophysique de Paris, 98 bis Bd
	  Arago, 75014 Paris, France
	  \and
          Universit\'e de Savoie, Chamb\'ery, 73011, France}

\date{Received 6 December 2002 / Accepted 27 March 2003}

\abstract{
In a companion paper, we investigated the question of the 
spatial origin of the cosmic rays detected in the Solar neighborhood, in 
the case of standard sources located in the Galactic disk.
There are some reasons to believe that there may also be a large 
number of sources located in the halo, for example if the Galactic 
dark matter is made of supersymmetric particles or if Primordial 
Black Holes are present. These exotic sources could enhance the 
$\bar{p}$, $\bar{d}$ or positrons above the standard background, 
indicating the existence of new physics.
The spatial distribution of these hypothetical sources, though an important 
ingredient to evaluate these exotic signals, is poorly known.
The aim of this paper is to point out that 
this discussion should not be disconnected from that of the
propagation properties in the Galaxy.
More precisely, we determine the regions of the halo from which a 
significant fraction $f$ of cosmic rays antiprotons and antideuterons
detected in the Solar neighborhood were emitted
(we refer to these regions as $f$-volumes),
for different sets of propagation parameters consistent with B/C data, 
as derived in \citet{Maurin02b}.
It is found that some of them lead to rather small
$f$-volumes, indicating that the exotic 
cosmic rays could have a local origin (in particular 
for a small halo or a large Galactic convective wind), coming from 
the solar neighborhood or the Galactic center region.
It is also found that the dark matter density enhancement (spike) due 
to the accretion around the central supermassive black hole gives a 
negligible contribution to the exotic charged particle signal on Earth.
The case of electrons and positrons is also discussed.
}

\maketitle


\section{Introduction}

A great amount of work has been done these last twenty years on the astrophysical 
signatures that could unravel new physics. In the eighties,
there were great hopes that the antiproton signal,
which showed an excess at an energy of a few hundreds 
of MeV in the first balloon experiments, could be such a signature. 
However, this hope was swept away by the progress in measurements 
-- see e.g. {\sc bess} \citep{Orito00,Maeno01}
or {\sc heat} \citep{Beach01}
and {\sc caprice} \citep{Boezio01}
at higher energy -- and a better 
determination of the cosmic ray propagation parameters (see e.g. \citet{Maurin02b}). 
It was shown that the measured
antiproton flux was indeed compatible with the sole secondary standard 
spallative production 
\citep{Bergstrom99b,Donato01}  (see the first 
paper for a comprehensive historical discussion and a panel of 
references dealing with exotic antiproton production).

\citet{Donato00} showed that the antideuteron
($\bar{d}\,$) signal could lead to a clearer signature of SUSY.
However, as discussed in many other studies on SUSY antiprotons 
\citep{Rudaz88,Stecker89,Jungman94,Bottino95,Bottino98,Wells99,Bergstrom99b}, 
the indeterminacy in the dark matter distribution, as well as its possible 
clumpiness \citep{Bergstrom99}, might severely change the 
conclusions. In contrast, the Hawking evaporation of Primordial Black
Holes (PBH) could also yield a new source of cosmic rays 
\citep{Maki96},
but the precise shape of the dark matter in this case is not crucial
\citep{Barrau02,Barrau03}. Nevertheless, in the latter case, it
was shown that even considering only the propagation parameters giving a
good fit to B/C data, 
the remaining degeneracy for example in the diffusive
halo height has sizeable effects on the primary flux \citep{Barrau02}.

Hence, at least two different phenomena can affect the conclusions
reached in papers dealing with exotic flux calculations. The first one,
related to the spatial distribution of SUSY sources, is usually
thoroughly discussed \citep{Bergstrom99b}, but the second point - namely the
influence of various propagation parameters - is generally skipped, due
to the simplicity of the propagation models used.
The aim of the paper is {\em not} to compare the predicted $\bar{p}$, $\bar{D}$
fluxes with observations for different series of models, but rather to point out
which characteristics of the models actually play a role, in order to give
some physical insights and milestones for studies specifically devoted to
exotic flux evaluations. 

We apply the method described in \citet{Taillet03} to determine
the volumes from which a fraction $f$ of cosmic rays
reaching the Solar neighborhood were emitted, or equivalently
the volumes that contribute to the fraction $f$ of the total flux detected
in the Solar neighborhood. These volumes will be 
referred to as the $f$-volumes throughout the paper.

We find that depending on the diffusion
parameters (evaluated from a systematic study of standard CR, \citet{Maurin02b})
as well as on the source spatial
distribution, the spatial origin of cosmic rays may be quite local,
the particles detected in the Solar neighborhood having mostly been
created a few kpc away from the Solar
neighborhood in some cases, or a few kpc away from the Galactic
center in others.


\section{Evaluation of the $f$-volumes}
\label{general_formula}

In a companion paper \citep{Taillet03}, 
we presented a method to compute the region from which a cosmic ray 
detected in the Solar neighborhood has a given probability of originating.
This method was applied to standard sources located in the disk, and 
we now use it for (exotic) sources in the halo.
A schematic view of our model is presented in Fig.~\ref{galaxy} where the 
isothermal dark matter profile has been superimposed on the Galaxy to 
compare their typical scales (the reader is referred 
to \citet{Taillet03} for all the details concerning the model, such 
as the functional form of the galactic wind and the geometry of the 
box).
The probability that a 
particle detected in the Solar neighborhood was emitted from any finite volume  
${\cal V}$ can be computed as
\begin{equation}
      {\cal P} \left\{ {\cal V} | \vec{r}_o \right\}=
     \frac{\int_{\cal V} w(\vec{r}_s) N_{\rm r_s}(\vec{r}_o) d^3\vec{r}_s}
     {\int_{\cal V_{\rm tot}} w(\vec{r}_s) N_{\rm r_s}(\vec{r}_o) d^3\vec{r}_s}\;\;,
     \label{proba_integree} 
\end{equation}
where the source distribution $w(\vec{r}_s)$ has been introduced and
$N_{\rm r_s}(\vec{r}_o)$ is the density in $\vec{r}_o$ resulting from 
a point source located in $\vec{r}_s$.
In this paper, we are interested in determining $f$-volumes, i.e.
volumes ${\cal V}(f)$ from which a given fraction $f$ of cosmic rays detected in the 
Solar neighborhood were emitted. They are defined by
\begin{equation}
      {\cal P} \left\{ {\cal V}(f) | \vec{r}_o \right\}= f\;.
      \label{def_f_vol}
\end{equation}
Actually, even for a given value of $f$, there are many different volumes, 
delimited by different closed surfaces, fulfilling this condition.
We focus on the smallest of these volumes, precisely delimited 
by an isodensity surface. 
Monte Carlo integration is then particularly well adapted to evaluate 
the integrals in Eq.~\ref{proba_integree}.
In a typical run, $\sim 10^6$ points are required to reach 
a $\lesssim 0.5$\% convergence and the integral is performed inside
all isodensity surfaces at once, so that
the $f$-volume defined by Eq.~(\ref{def_f_vol}) are simple to recover.

\subsection{Influence of the propagation parameters}
\label{PL-sources}

\begin{figure}[hbt!]
\centerline{
\includegraphics*[width=\columnwidth]{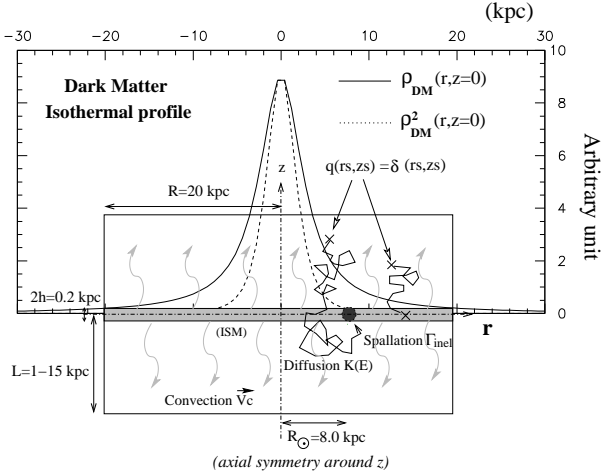}}
\caption{Schematic view of our Galaxy: diffusive and convective propagation
plus spallations in the thin disk. 
Effective primary exotic sources follow either the dark matter profile
or its square (isothermal profile is depicted).}
\label{galaxy}
\end{figure}

The quantity $N_{\rm r_s}(\vec{r}_o)$ appearing in 
Eq.~(\ref{proba_integree}) is evaluated by solving the 
diffusion equation with a point-like source, in the geometry depicted 
in Fig.~\ref{galaxy}.
Propagation is affected, at different levels, by three effects: escape, galactic wind and 
spallations.
First, escape happens when a particle reaches one of the boundaries of the 
diffusive volume. As discussed in the companion paper, this limits 
the range from which cosmic rays can travel to the Solar neighborhood.
It was also shown that the side boundary plays only a minor role,
and one can assume that the box has an infinite radial extension.
Second, a convective wind $V_c$ directed out from the 
Galactic plane blows the charged nuclei away, so that it is more
difficult to reach the plane from high $z$ sources. 
Finally, spallations may happen when a nucleus crossing the thin disk interacts with 
the interstellar matter.
The nuclei are then destructed at a rate $\Gamma_{\rm inel}=2hn_{\rm ISM}.v.\sigma_{\rm inel}$.
A particle emitted from a remote source is more affected by 
spallations as it is likely to have crossed the disk many times before 
reaching the Solar neighborhood.
In the companion paper, this effect was shown to be important for heavy 
species created in the disk. Here, we focus on very light species, 
having smaller cross-sections, which are mostly created in the halo.
They are affected by the wind in the whole halo, i.e. 
from the moment of their creation, whereas they are only affected by spallations 
when they cross the disk, which is less likely for halo sources than 
for disk sources.
As a result, spallations play only a minor role in the present study
(this effect is nevertheless included in our treatment).

When these three effects are taken into account, the density in $O$ due to a Dirac source 
$\delta(\vec{r}-\vec{r}_s)$ can be computed. 
Because of the cylindrical symmetry present for an infinite disk, 
it is equivalent to consider a source term $\delta(z-z_s) 
\delta(r-r_s)/2\pi r_s$, which leads to
\begin{equation}
    N_{(r_s,z_s)}(0)=e^{-z_s/r_{\rm w}}
    \sum_{i=1}^{\infty}
    \frac{J_0\left(\zeta_i r_s/R \right)}{\pi J_1^2(\zeta_i)R^2A_i}\times
    \frac{\sinh\left[S_i(L-z_s)/2\right]}{\sinh (S_iL/2)} 
    \label{sol_vent}
\end{equation}
with
\begin{equation}
     S_i = \sqrt{\frac{4}{r_{\rm w}}+\frac{4\zeta_i^2}{R^2}}
     \;\; \mbox{and} \;\;
     A_i = K\left( \frac{2}{r_{\rm sp}} +\frac{2}{r_{\rm w}} +
     S_i \coth \left[\frac{S_iL}{2} \right]\right)
     \label{ai_reduit}
\end{equation}
and where the parameters
\begin{equation}
    \left\{
    \begin{array}{l}
	\displaystyle
	r_{\rm w}  \equiv \frac{2K}{V_c} \approx
	5.87 \unit{kpc} \times \frac{K(E)}{0.03
	\unit{kpc}^2\unit{Myr}^{-1}}
	\; \frac{ 10 \unit{km} \unit{s}^{-1}}{V_c},
	\label{f_rw}\vspace{0.2cm}\\\displaystyle
	r_{\rm sp}\equiv \frac{2K}{2h \Gamma_{\rm inel}} \approx
	\frac{3.17 \unit{kpc}}{\beta} \times\frac{K(E)}{0.03
	\unit{kpc}^2\unit{Myr}^{-1}}
	\; \frac{ 100 \unit{mb}}{\sigma},
	\label{f_rspal}
    \end{array}
    \right.
\end{equation}
give the order of magnitude of the typical distance over which the associated process 
affects propagation. 
\begin{figure}[hbt!]
    \centerline{
    \includegraphics*[width=\columnwidth]{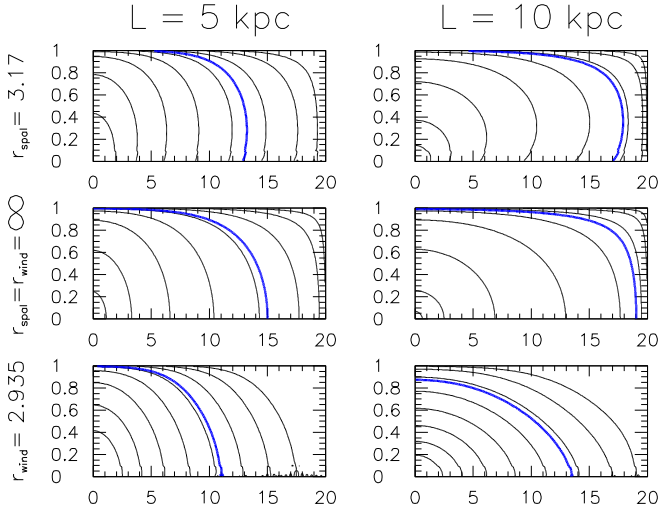}}
    \caption{Isodensity surfaces in the $(z,r)$ plane for $L=5$~kpc and $L=10$~kpc 
    (side boundary $R=20$~kpc).
    Inner contours correspond to $d{\cal P}(r_s,z_s|O)/d^3\vec{r}_s=0.01 \, 
    \mbox{kpc}^{-3}$  and the contours are spaced by a factor $1/4$. From top to 
    bottom, $r_{\rm sp}=3.17$~kpc (no wind), no wind and no spallations, 
    $r_{\rm w}=2.935$~kpc (no spallations).
    These numbers correspond, respectively, for a reasonable choice of $K(E)$
    at 1~GeV, to  $\sigma_{\rm sp}\approx100$~mb
    and $V_c\approx20$~kpc~Myr$^{-1}$. The additional thick line in each panel 
    delimitates contours ${\cal P}({\cal V}(f)|O)=99\%$.}
    \label{fig:prim_wind_spal}
\end{figure}
In practice, large values of $R$ have been used in Eq.~(\ref{sol_vent}) so that the 
hypothesis $R\rightarrow \infty$ is actually recovered.
The effect of escape, wind and spallations are compared in Fig.~\ref{fig:prim_wind_spal}, which 
shows the shape of the isodensity surfaces for two values $L=5$kpc 
(left panels) and $L=10$~kpc (right panels), and for typical values 
of $r_{\rm w}$ and $r_{\rm sp}$. 
The value $r_{\rm sp}=3.17$~kpc has been retained because it corresponds
to the antideuteron destruction cross section for a typical value of the diffusion coefficient
$K=0.03$~kpc$^2$~Myr$^{-1}$. 
In the upper panels, one can see the shrinking of the contours in the vicinity 
of the disk, due to the effect of spallations.
The effect of the wind is rather to flatten the contours, as can be 
seen in the lower panels. 
The probability density also decreases more rapidly when convection or spallations 
are included than when diffusion alone is considered. 
As a result, the 99\%-volumes are reduced, as indicated by the thick lines. 
It is thus of importance to use realistic values for $K(E)$, $L$ and $V_c$ 
in order to give confident $f$-volumes for real situations ($\sigma_{\rm inel}$
is not a free parameter, it solely depends on the species we consider).

To summarize the previous results about the origin of exotic primaries
in diffusion/convection/spallation models: 
i) the pure diffusive regime provides an upper limit that is
strongly dependent on the halo size;
ii) the Galactic wind lessens the $f$-volumes:
 either propagation is convection-dominated -- in this case, the origin
depends only on the value of $L$ and $r_{\rm w}$, i.e. $V_c$ and
$K(E)$ -- or it is escape-dominated and the geometrical upper limit 
(sole dependence on $L$, not $K$) is recovered;
iii) spallations also systematically lessen the $f$-volumes:
the heavier the nucleus, the larger its destruction rate, the closer 
it comes from. However, as a particle created in the halo is less likely 
to cross the disk, this effect is negligible compared to the wind for
$r_{\rm w}\gtrsim r_{\rm sp}$.
We show below that  all these effects are more pronounced for annihilating SUSY 
than for evaporating PBH because the density profile $h_{\rm DM}(r,z)$ 
appears with a square.

\subsection{Dark matter distribution}

The dark matter distribution in our Galaxy is poorly known, and 
several dark matter profiles can be used. 
The first constraint is that the observed rotation curve of our Galaxy is
almost flat beyond a few kpc from the center.
For a spherical halo, it follows that the density decreases
as $1/r^2$ outside the central regions.
In the inner regions, the situation is far from clear.
Numerical simulations indicate that the central distribution of dark
matter is cuspy, with a $r^{-\gamma}$ dependence with $\gamma \sim 0.5 -
1.5$ \citep{Ghez98},
but this seems to be in contradiction with observations
\citep{Binney01}.

In the absence of a clear answer to this problem, we use
several profiles for the Dark Matter distribution, with the
generic form
\begin{equation}
    h_{\rm DM}(r,z) = \left(
    \frac{R_\odot}{\sqrt{r^2+z^2}}
    \right)^\gamma
    \left( \frac{R_c^\alpha + R_\odot^\alpha}{R_c^\alpha +
    (\sqrt{r^2+z^2})^\alpha}
    \right)^\epsilon
    \label{dep_spatiale}
\end{equation}
where spherical symmetry has been assumed.
Numerical simulations point toward singular profiles with
$\gamma = 1.5$, $\alpha = 1.5$, $\epsilon = 1$ and $R_c=33.2
\, \mbox{kpc}$ \citep{Moore99} or
$\gamma = 1$, $\alpha = 1$,  $\epsilon = 2$ and $R_c=27.7
\, \mbox{kpc}$ (\citep{NFW}, hereafter NFW).
We also considered an isothermal profile with $\gamma = 0$, $\alpha =
2$, $\epsilon = 1$ and $R_c=3$~kpc (the modified isothermal profile
would give very similar results).

As already said, exotic SUSY particles (resp. PBH) are supposed to 
fill (resp. follow) the dark matter halo profile $h_{\rm DM}(r,z)$. 
However, the nature of the cosmic ray creation process is different in 
these two cases, leading to very different effective source terms, i.e. different weight 
$w(r,z)$ in Eq.~(\ref{proba_integree}). 
For evaporating Primordial Black Holes, the particle production is 
proportional to the density of the objects $w_{\rm PBH}(r,z) \propto 
h_{\rm DM}(r,z)$.
In contrast, the production term for supersymmetric  particles 
is proportional to the square of 
the density because two dark matter particles must be present for 
annihilation to occur. 
In this case $w_{\rm SUSY}(r,z) \propto  h_{\rm DM}(r,z)^2$.

\begin{figure}[hbt!]
\centerline{
\includegraphics*[width=\columnwidth]{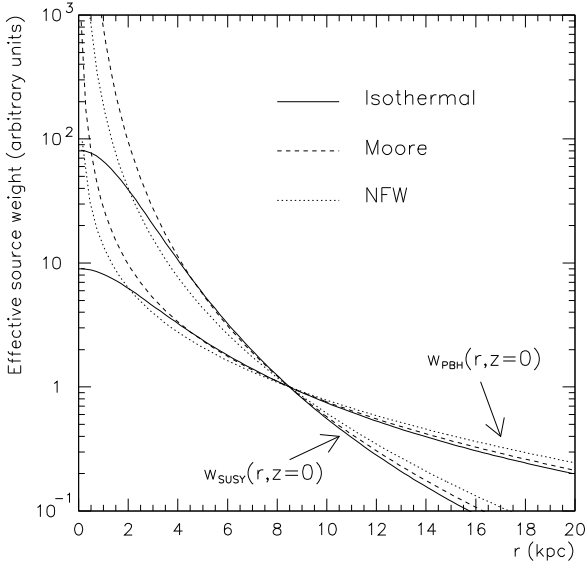}}
\caption{Effective source weight (PBH or SUSY) for several profiles
(see text).}
\label{profils}
\end{figure}

They are displayed in Fig.~\ref{profils} both for SUSY and PBH weight
(see above).
The Moore and NFW profiles are singular at the Galactic center, so that 
the source term is much stronger there. The probability that a cosmic 
ray detected in the Solar neighborhood was emitted from this region is enhanced for 
these profiles.
A crude estimate of this effect is obtained by a mere count of
the effective (PBH or SUSY) source numbers in this critical region.
For example, in the range [$0-2$]~kpc, a Moore profile
leads to an enhancement $\times 2.7$ for PBHs and $\times 90$ for SUSY annihilations,
compared to the isothermal case.
Stretching this interval decreases the enhancement factor, and for [$0-4$]~kpc,
it is respectively $\times 1.5$ and $\times 25$, and finally for
[$0-8$]~kpc, the numbers are $\times 1.1$ and $\times 20$.
The enhancement is far smaller for PBH than for SUSY particles.
Notice that the upper limits on the PBH density derived from  antiproton 
flux measurements in \citet{Barrau02} were of the same order of 
magnitude for an isothermal halo and for cuspy halos. 
This result is definitively not transposable to the SUSY case.

This is not the final word. The center of our Galaxy contains a supermassive 
black hole (SBH) of a few $10^6$ M$_\odot$. During its formation, 
it probably accreted the surrounding dark matter, leading to a 
local enhancement of the density.
Gondolo \& Silk (1999) (hereafter GS) found that if the SBH grows adiabatically 
in the center of the Galaxy, the cuspy profile ($\rho(r)\propto r^{-\gamma}$ 
with $0<\gamma<2$) becomes spiky and $\rho(r)\propto 
r^{-A}$ with $2.25<A<2.5$ in a region of a few parsecs around the black hole.
The presence of the spike would have dramatic consequences
for several predictions of the signal from annihilating dark matter particles,
e.g. $\gamma$ and neutrinos \citep{Gondolo_Silk99} or synchrotron emission
of $e^+e^-$ pairs \citep{Gondolo00,Bertone01}.
The signal coming from the direction of the Galactic Center is 
obtained by integrating along the line of sight, and the contribution 
of the central region is very different with or without a spike. 
In the case of the isothermal profile, the central region (around the SBH)
contributes at the level of $\sim 10^{-9}$ whereas this contribution is greater
than $\sim 10^5$ for a Moore profile \citep{Gondolo_Silk99}.
However, these results are expected to be overoptimistic,
and it is doubtful that such a spike exists in our Galaxy, as 
indicated by a more careful dynamical modelling of the SBH growth
\citep{Ullio01}.
These authors review several effects (adiabatic growth versus instantaneous growth,
models with off-centered black holes) and recover
some results that were known before the Gondolo \& Silk paper:
only the peculiar case in which the SBH forms adiabatically at the exact
center of the dark matter profile can lead to  an enhancement such as described in GS. 
Finally, in a recent study, \citet{Merritt02} have observed that, taking into account 
the quite large probability that the Milky Way experienced a major merger 
in its history, the ensuing dark matter profile and
resulting annihilation fluxes could be several order
of magnitudes smaller than obtained with dark matter profile
not disturbed by a SBH.

The points discussed above are mostly relevant for particles
travelling in straight lines. For charged particles,
due to the diffusive nature of propagation, the probability to come
from a sphere ${\cal S}$ of radius $r=10$~pc around on the Galactic center 
($\sim  8$~kpc away) is $\int_{\cal S}(d{\cal P}/d^3\vec{r}_s)d^3\vec{r}_s$, which 
is $\lesssim 10^{-10}$ ($d{\cal P}/d^3\vec{r}_s$ is given for example in
Fig.~\ref{fig:prim_wind_spal}).
Due to the very narrow scale where the SBH may affect the distribution,
even enhancement such as obtained in \citet{Gondolo_Silk99} -- and which is not very realistic
--  cannot yield a significant contribution for charged particles. 
Eventually, the dark matter profile remains of importance (isothermal or cuspy).
In the following, most results will be presented for the isothermal 
case, the influence of the cusp being discussed at the end.

\subsection{$f$-volumes for SUSY and PBH weights and different values 
of $L$}
\label{susy_et_les_jaunes}
			
We now have all the elements to compute the $f$-volumes, inserting the 
source distributions described above in Eq.~(\ref{proba_integree}).
The function entering the integral does not possess cylindrical 
symmetry, so that the full three-dimensional 
integral must be computed.  
We first neglect spallations and galactic wind to consider only the effect of $L$.
This parameter is expected to play an important role, 
as the charged particles created outside of the magnetic halo of our 
Galaxy do not penetrate inside it and are not detected 
\citep{Barrau02,Barrau03}.
\begin{figure}[hbt!]
    \centerline{
    \includegraphics*[width=0.5\columnwidth]{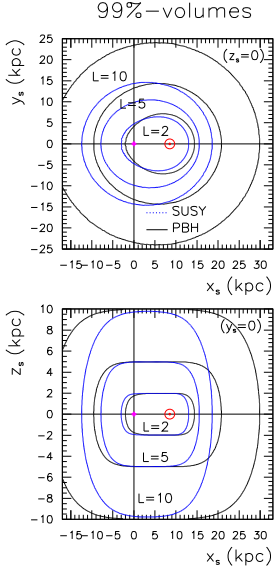}}
    \caption{Contours ${\cal P}({\cal V_{\rm SUSY,~PBH}}|R_\odot)=99\%$ 
    origin for $L=2$, $L=5$ and $L=10$~kpc. $f$-volumes have been
    respectively evaluated with weight $w_{\rm PBH}(r,z)\propto h_{\rm DM}(r,z)$ 
    (solid lines) and $w_{\rm SUSY}(r,z)\propto h^2_{\rm DM}(r,z)$ (dashed lines)
    for $R\gg L$, but the result remain
    mostly unchanged using $R=20$~kpc (but in case of a halo size $L=10$~kpc 
    that requires $R=30$~kpc).
    Upper panel: $V_{\rm SUSY}(99\%)$ and $V_{\rm PBH}(99\%)$
    in the $x_s-y_s$ plane ($z_s=0$). Lower panel: same quantities
    but in the $x_s-z_s$ plane ($y_s=0$).
    In both panels, the dot marks the Galactic center, and $\odot$
    denotes the Sun location (it is set to $R_\odot=8$~kpc).}
    \label{proba_vol_weighted}
\end{figure}
Fig.~\ref{proba_vol_weighted} shows the $99$\%-volumes 
in the Galactic plane $z_s=0$ (upper panel), for the PBH and SUSY case. 
Their shape reflects the fact that
the probability density has a maximum at $r_s=R_\odot$, while the effective source 
distribution peaks at the Galactic center $O$. 
Because of the quadratic dependence on $h_{\rm DM}(r,z)$, 
${\cal V}_{\rm SUSY}(99$\%) is smaller than ${\cal V}_{\rm PBH}(99$\%). 
Three halo sizes are displayed ($L=2$, $5$ and $10$~kpc): for larger halos,
the surfaces are more deformed towards the Galactic center (the contribution 
of this region to the flux is larger), whereas they remain grossly unaffected
in the anti-center direction. This effect is less pronounced in the
case of a PBH-like source distribution. The same contours are also 
plotted for $y_s=0$ in the lower panel. 
The shapes are almost maximally distorted towards rectangular contours. 
This is less and less pronounced, as either $L$ is enhanced, or larger 
powers of $h_{\rm DM}(r,z)$ are chosen.

The figures above show clearly that we are only sensitive to 
a well-defined region of the source distribution: first to the region which is 
embedded in the diffusive halo, and then, even within this region, to a 
sub-region between the Galactic center and the Solar neighborhood.
These sub-regions represent a fraction of the total number of sources 
given by
\begin{displaymath}
    f^{\rm tot}(L) = \frac{\int_{{\cal V}(99\%)} w(r,z)\; 
    d^3\vec{r}}{\int  w(r,z)\; d^3\vec{r}}\;,
\end{displaymath}
where $\int  w(r,z)\; d^3\vec{r}$ is the total number of sources.
It is also of interest to compare the number of sources located 
in the same sub-regions to the number of sources in the diffusive halo
\begin{displaymath}
    f^{\rm cyl}(L) = \frac{\int_{{\cal V}(99\%)} w(r,z)\; 
    d^3\vec{r}}{\int_{\cal V_{\rm cyl}} w(r,z)\; d^3\vec{r}}\;;
\end{displaymath}
where ${\cal V}_{\rm cyl}$ is the volume of the diffusive halo.
The corresponding numbers are given in Table~\ref{tab1} 
for various halo sizes.
\begin{table}[ht]
\begin{center}
\begin{tabular}{|c||c|c|}   \hline
       & $f^{\rm cyl}(L)$ & $f^{\rm tot}(L)$ \\
        & PBH~/~SUSY  & PBH~/~SUSY\\\hline\hline
$L=10$~kpc & $\sim 1.$~~~~~~$\sim 1.$     & 0.023~~~~~~0.76\\\hline
$L=5$~kpc  & 0.70~~~~~~0.85 	 	& 0.010~~~~~~0.54\\\hline
$L=2$~kpc  & 0.31~~~~~~0.60		& 0.002~~~~~~0.21 \\\hline
    \end{tabular}
    \caption{Fraction of the number of exotic primaries emitted in 
    $V_{\rm PBH}(99\%)$ and $V_{\rm SUSY}(99\%)$ for various $L$, 
    compared to the total number of exotic primaries emitted
    either in the bounded geometry (halo size $L$ and radial 
    extension $R=20$~kpc) -- denoted $f^{\rm cyl}(L)$ --, or in 
    the whole dark halo -- denoted $f^{\rm tot}(L)$.}
  \label{tab1}
\end{center}
\end{table}
The fraction $f^{\rm cyl}(L)$ decreases with $L$, much faster
for PBH than for SUSY. This can be understood as the number of contributors
outside of the dark halo core radius rapidly vanishes for SUSY particles
(see Fig.~\ref{galaxy}).
As regards the results for $f^{\rm tot}(L)$, we remark that this number 
is particularly small for PBH, i.e. only a very small fraction of
primordial halos distributed
in the Galaxy contribute to the charged primary cosmic rays detected 
in the Solar neighborhood.

Finally, it is also interesting to give the fraction of 
primaries that escape before reaching the Solar neighborhood.  
It is defined as
\begin{equation}
  f^{\rm esc}({\cal V})\equiv  1 - 
  f^{\rm detect}({\cal V}) 
  = 1- \frac{\int_{{\cal V}} w(r,z)
     	N_{\rm cyl}(\vec{r}|R_\odot)\;d^3\vec{r}}{
	\int_{{\cal V}} w(r,z)
     	N_{\infty}(\vec{r}|R_\odot)\;d^3\vec{r}} \;\;,
     \label{rir_SN}
\end{equation}
where $N_{\rm cyl}(\vec{r}|R_\odot)$ and $N_{\infty}(\vec{r}|R_\odot)$
are respectively related to the flux of particles detected at $R_\odot$,
in the cylindrical geometry and in an unbounded space, from the same
sources
emitting from inside the volume ${\cal V}$. Estimations for $L=10$~kpc and 
$L=2$~kpc are compiled in Tab.~\ref{tab2}. 
\begin{table}[ht]
    \begin{center}
	\begin{tabular}{|c||c|c|}   \hline
	    & $L=10$ kpc & $L=2$ kpc \\\hline\hline
	    $f^{\rm esc}_{\rm PBH}:~{\cal V}$(50-90-99\%) & 40-55-64\%&
	    45-75-88\%\\\hline
	    $f^{\rm esc}_{\rm SUSY}:~{\cal V}$(50-90-99\%)& 49-52-55\%  &
	    59-92-95\%\\\hline
	\end{tabular}
	\caption{Fraction of primaries $f^{\rm esc}$ 
	emitted from the (50-90-99)\%-volumes of the cylindrical geometry (see
	above)
	that escape through upper and lower boundary located at $L=10$~kpc or
	$L=2$~kpc (for PBH and SUSY effective source distribution), before 
	they can reach the Solar neighborhood.}
	\label{tab2}
    \end{center}
\end{table}
The trends are conform to intuition. Forming greater fractions of the
detected flux requires more distant sources, the latter more easily escape
through boundaries. For large diffusive halo $L$, the 
fraction that escape increases more quickly for PBH sources than for SUSY 
sources, whereas the converse is true for small halos. 
This is related to the fact
that one has to compare the shape and typical extension of
the source distribution to the parameter $L$.
The fraction of primaries which are emitted inside 
${\cal V}$ but which never reach the solar neighborhood is actually 
greater than $f^{\rm esc}$, as even in the case of diffusion in 
unbounded space, there are many trajectories which start in ${\cal V}$ 
and never  reach the solar neighborhood (diffusion in three dimensions 
is a transient process).


\section{Realistic propagation parameters}
\label{BC_induced}
The previous section considered simplified diffusion situations
with a typical value $K\sim0.03$~kpc$^2$~Myr$^{-1}$.
Actually, $K(E)$ is energy dependent, and more precisely,              
\begin{displaymath}
    K(E)=K_0\beta {\cal R}^\delta.
\end{displaymath}
Here, $\delta$ is the diffusion slope and $K_0$ the normalization
of the diffusion coefficient.
In a previous study (see Paper Ia, Ib), we show that various combinations 
of parameters $K_0$, $\delta$, diffusive halo height $L$ and Galactic wind
magnitude $V_c$ are equivalent, in the 
sense that they give a B/C spectrum that is consistent with the 
observations.
In this section, we use these combinations to provide a realistic 
range of values for $r_{\rm w}$ and $r_{\rm sp}$ and to explore
the consequences on the origin of exotic primary antiprotons 
and antideuterons. 
The heavier antinuclei will not be considered here, as it was shown by
Chardonnet, Orloff \& Salati (1997) that their formation is suppressed 
because of the low probability of coalescence of many antinucleons.

			 
\subsection{Parameter range allowed}
To compute the  parameters introduced in Eqs.~(\ref{f_rspal}), the 
spallation cross sections of antiprotons and antideuterons are taken 
from the Particle Data Group\footnote{http://pdg.lbl.gov/}. In this work, 
we only consider spallation on pure 
hydrogen. It would be straightforward to take into account the 
spallations on the Helium component of the interstellar medium, but 
the effect is too small to be worth the complication.
The four parameters $K_0$, $\delta$, $L$ and $V_c$ are taken from 
our comprehensive study of standard secondary to primary B/C ratio
\citep{Maurin02b}. 
Three values (two extremes and a medium value) have been retained 
for both the diffusion slopes 
($\delta=0.35$, 0.60 and 0.85) and the halo sizes 
($L=2$~kpc, $L=6$~kpc and $L=10$~kpc). 
We emphasize that the values of all these parameters come from the 
study of standard sources of cosmic rays and do not depend on the 
exotic sources, which do not produce B nor C.
We do not take reacceleration and energy losses into account in this 
work. These effects, though necessary to study the spectra of cosmic 
rays, are not so crucial here as they only amount to a redistribution of 
the cosmic rays at different energies. 
A particle detected at an energy of 1 GeV/nuc was just created at a slightly 
different energy and its origin is not drastically different.

\begin{figure}[hbt!]
\centerline{
\includegraphics*[width=.85\columnwidth]{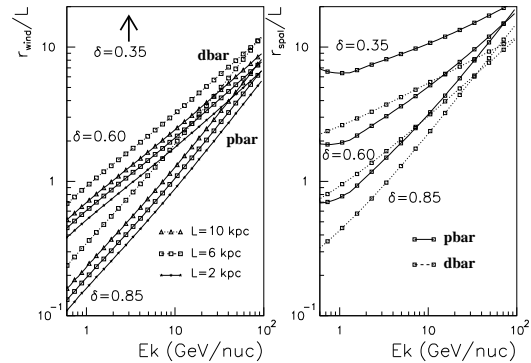}}
\caption{Left panel: evolution of $r_{\rm w}$ as a function of kinetic energy per
nucleus for primary $\bar{p}$ and $\bar{d}$; from top to bottom, 
$\delta=0.35$, $\delta=0.60$ and $\delta=0.85$. The parameter 
$r_{\rm w}/L\equiv\chi_{\rm w}$, as well as $r_{\rm sp}/L\equiv\chi_{\rm sp}$, 
are not very sensitive to the halo size $L$ (for $\bar{d}$, only $L=6$~kpc
is displayed) but $r_{\rm w}$ and $r_{\rm sp}$ do. Right panel: 
$r_{\rm sp}/L$ as a function of $E_k$/nuc for the same $\delta$ values
and for the halo size $L=6$~kpc.
The values of $r_{\rm w}$ are different 
between $\bar{p}$ and $\bar{d}$ because they depend on rigidity 
(through $K$), i.e. on $Z/A$ (it is $1$ for $\bar{p}$ and $1/2$ for $\bar{d}$). 
For $r_{\rm sp}$, there is an additional strong dependence on the species 
because of the destruction cross sections.}
\label{fig:rw_rs}
\end{figure}

The values of $r_{\rm sp}$ and $r_{\rm w}$ are plotted in 
Fig.~\ref{fig:rw_rs} for antiprotons and antideuterons. 
The left panel shows that
propagation is convection-dominated ($r_{\rm w}\ll 1$) at low energy 
when large $\delta$ values are considered and escape-dominated at 
all energies for small $\delta$.
Notice that although at a given $\delta$, the quantity $r_{\rm w}/L$ 
is fairly independent of $L$, the origin is definitely not the same 
for $L=2$ kpc as for $L=10$ kpc.
The right panel shows that spallation is not the dominant effect for the
light nuclei considered here. 
Only for large diffusion slopes $\delta$ and more particularly
for antideuterons this effect becomes 
sizeable and comparable to the diffusive escape. 
The comparison of the two panels shows that spallations are always 
less efficient than convective wind or boundary escape.
Finally, whatever the value of $\delta$, propagation is escape-dominated 
above a few tens of GeV/nuc and the origin of primary cosmic rays is 
solely dependent on the halo size. 


\subsection{Antiprotons and antideuterons}
\label{pbar_dbar_realiste}
We are now able to draw the $f$-volumes
for the realistic propagation parameters being considered.
We focus on the antideuteron signal as it seems to be the most promising
species to look for in cosmic rays. 
An interstellar energy of 1~GeV/nuc is chosen; the nuclei that reach the detector 
are solar modulated so that they are detected with a final energy of $400-800$~MeV/nuc, 
where the signal is the more interesting. 
Table~\ref{tab3} summarizes the values
of $r_{\rm w}$ and $r_{\rm sp}$ at this energy for antideuterons.
\begin{table}[ht]
    \begin{center}
	\begin{tabular}{|c||c|ccc|}   \hline
	    &(kpc) &$\delta=0.35$	& $\delta=0.60$	& $\delta=0.85$\\\hline\hline
	    $L=10$~kpc     & $r_{\rm w}=$       & $\infty$     & 8.  & 2.9\\
	    & $r_{\rm sp}=$       & 21.         & 7.6  & 3.5\\\hline
	    $L=6$~kpc	& $r_{\rm w}=$      & $\infty$     & 5.5  & 2.1\\
	    & $r_{\rm sp}=$       & 15.5        & 5.5  & 2.6\\\hline
	    $L=2$~kpc	& $r_{\rm w}=$      & $\infty$     & 2.1  & 0.85\\
	    & $r_{\rm sp}=$      & 6.	     & 2.2  & 1.05 \\\hline
	\end{tabular}
	\caption{$r_{\rm w}$ and $r_{\rm sp}$ for three halo sizes $L$ and 
	three diffusion slopes $\delta$: these numbers are for 1~GeV/nuc
	(interstellar energy) antideuterons.}
	\label{tab3}
    \end{center}
\end{table}
The situation is very different for small or large $\delta$.
For small values (corresponding roughly to a Kolmogorov power 
spectrum $\delta = 1/3$), only spallations affect the propagation
($V_c=0$, $r_{\rm w}=\infty$)
and this effect was shown to be weak; for large $\delta$  -- 
the value $\delta=0.85$ is the one preferred in our B/C analysis (Paper~Ib) 
--, models are convection-dominated though $r_{\rm sp}$ and $r_{\rm w}$
have about the same strength.

\begin{figure}[hbt!]
\centerline{
\includegraphics*[width=0.85\columnwidth]{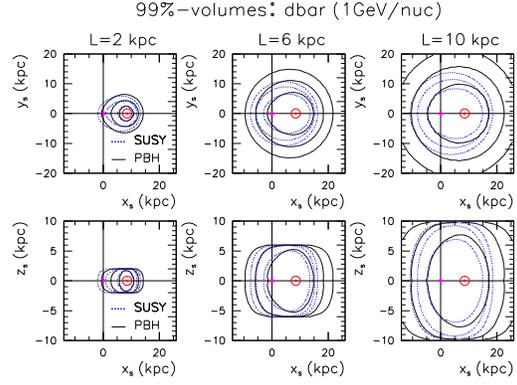}}
\caption{99\%-volumes for exotic primaries (no side boundaries). 
Upper panels: cut in the $z_s=0$~kpc plane; lower panels: cut in the 
$y_s=0$~kpc plane. 
Left panels correspond to $L=2$~kpc, middle panels to $L=6$~kpc and right panels to $L=10$~kpc. 
In each panel, we plot 
either the PBH case (solid lines) or the SUSY case (dotted lines).
From external lines to internal lines correspond the values of the diffusion
coefficient slope $\delta=0.35$, $\delta=0.60$, $\delta=0.85$.}
\label{fig:final1}
\end{figure}

Fig.~\ref{fig:final1} displays ${\cal P}_{\rm cyl}({\cal V}(f)|O)=99$\% 
for the values reported in Tab.~\ref{tab3}. For $\delta=0.35$ (external 
contours), the {\em geometrical\/} (upper limit) contours are recovered. 
However, for larger $\delta$ (internal contours), these contours shrink.
All comments made in Fig.~\ref{proba_vol_weighted} as regards halo 
size, or SUSY and PBH behavior, remain valid. 
Actually, the diffusion coefficient slope $\delta$,
as $L$ for the geometrical limit, is a key parameter to trace back
the CR origin, because of the values of $r_{\rm w}$ it implies, 
through $V_c$ and $K_0$. 

\begin{figure}[hbt!]
\centerline{
\includegraphics*[width=1.\columnwidth]{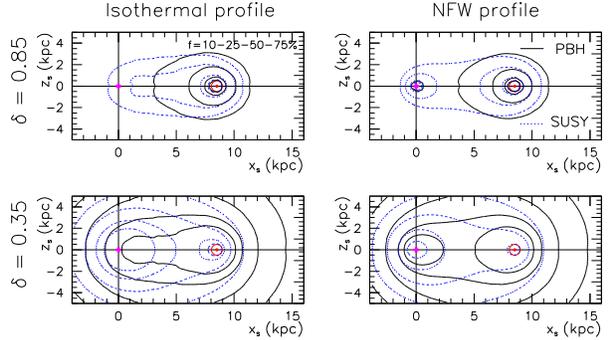}}
\caption{${\cal P}_{\rm cyl}({\cal V}(f)|O)=10\%-25\%-50\%-75\%$
for $L=10$~kpc in the $z_s=0$ plane (except for the $75$\%-volumes,
other $f$-volumes with $f\lesssim50$\% are not deformed by the boundaries so that
they present symmetry around the $x_s$ axis). Upper panels correspond 
to $\delta=0.85$, and lower panels to $\delta=0.35$. 
Both the PBH case (solid lines) and the SUSY case (dotted lines) are plotted.}
\label{fig:final3}
\end{figure}

It is also of interest to have a closer look at the first \% that contribute
to the flux. As the $f$-volumes with $f\lesssim50$\% correspond
to isodensity contours that are quite insensitive to the boundaries (or 
to other effects) they present the axial symmetry around the $x_s$ axis, 
so that a single cut through, e.g. the $x_s-z_s$ plane, delivers all the 
information about their shape.
Fig.~\ref{fig:final3} displays  the $f$-volumes $f=$10-25-50-75\%
 for $L=10$~kpc.
The difference observed in Fig.~\ref{fig:final3} between small (lower panels) 
and large $\delta$ (upper panels) is readily explained: a large value of $\delta$ also 
corresponds to a large value of $K_0$ (see \citet{Maurin03a} for details),
so that a greater wind is needed in order to prevent from too many spallations 
occurring at low energy. 
The net result is that the wind blow  the particles away, 
reducing the effective zone from where they come from. 
This is not the case for small $\delta$ where the {\em geometrical\/} limit 
(pure diffusion) is almost reached. 
The consequences for indirect dark matter searches are important. 
In the case of an isothermal profile (left panels), 
the particles created in the Galactic center have 
a small probability to reach the detector for large $\delta$, whereas 
the converse is true for small $\delta$.
In the latter case, the predictions and the limits that can be put on 
a supersymmetric signal depend heavily on the 
central shape of the dark matter halo, which is precisely the part we 
know the least about.
These contours for smaller diffusive halo sizes $L$ have not
been presented; they have a smaller extent, meaning that we are less 
sensitive to the distribution of dark matter far from the Solar 
neighborhood. 
As a result, the question of the dark matter density profile cusp is
less crucial for small $L$.

Similar contours for the NWF profile are drawn in the right panels
of Fig.~\ref{fig:final3}. 
Combining information from the above surfaces to the relative
enhancement of sources going from the isothermal case to the cuspy case
allows several complementary remarks: for small $\delta$,
about half the SUSY Cosmic rays come from the range [$0-3$]~kpc. 
Thus, the $\sim 50$ enhancement factor on the production provided by 
the cusp translates directly into a factor $\times 50$ in the 
detected flux.
For PBH case, the origin is less localized and the enhancement factor 
is smaller, so that the net gain is more probably about $10-20$\%.
For large $\delta$, contours look like boxes encompassing both the Solar position
and the Galactic center. In the SUSY case, the addition of a cusp strongly 
deforms the box towards the Galactic center, but it is not 
straightforward  to estimate the enhancement without considering 
specific values for the diffusion parameters. 
For PBH, the contours, and hence the flux, are not expected to be very 
sensitive to the parameters. 
This discussion is of less importance for small halo sizes.

From the above discussion, it appears that the most important 
parameters are $L$ and $V_c/2K$. The value $\delta = 1/3$  
(Kolmogorov spectrum) corresponding to  $V_c = 0$, 
has been preferred these last years (see e.g. \citet{Strong98}).
However, our previous studies 
\citep{Maurin01,Maurin02b,Maurin03a}
show that large values of $\delta$, and non-null values of  $V_c $, are preferred. 
This trend is confirmed the most recent results of 
\citet{Moskalenko02} who now tend to prefer $\delta=0.42-0.52$. 
To conclude, if the value of $\delta$ happens to be large 
or more precisely if a strong Galactic wind is preferred, the discussions 
about the dark matter profile, including about the existence of a spike,  
are not so crucial. 
If conversely $\delta$ is small (no wind), the dark matter cusp
as well as the exact location of the Solar system should be accurately
known before exploring the SUSY parameter space.

Finally, all the remarks made for antideuterons in the previous sections 
apply as well for antiprotons. According to
Fig.~\ref{fig:rw_rs}, for a given $\delta$ at a given energy, the 
corresponding $r_{\rm w}$ is about twice its antideuteron value. 
The resulting $f$-volumes are larger than those for antideuterons, but 
the conclusions remain the same.

			 
\subsection{Electrons and positrons}
\label{subsec:electrons_positrons}
Exotic sources in the halo also emit electrons and positrons.
Positrons are more promising to study supersymmetric signals as the 
background of standard positrons is much lower than electron's 
($e^+/(e^++e^-)<0.1$), being predominantly secondary.
Recently, the {\sc heat} experiment \citep{Coutu99} reported an excess
at about 7~GeV (see also the {\sc mass}-91 experiment, \citet{Grimani02}).

These particles are lighter than nuclei, so that they are subject to much 
stronger energy losses, due to synchrotron radiation and inverse Compton. 
This results in an effective lifetime given 
by \citep{Aharonian95,Atoyan95,Baltz99}
\begin{displaymath}
      \tau_{\rm loss} \sim 300 \unit{Myr} \times \frac{1 \unit{GeV}}{E} \,.
\end{displaymath}
\citet{Aharonian95} and \citet{Atoyan95} showed that in that case, all boundaries
have negligible effects on positrons and electrons above a few GeV,
so that the characteristic distance travelled by these species is 
$r_{\rm loss} \sim \sqrt{K \tau_{\rm loss}}$ (random walk through the tangled 
magnetic fields), or
\begin{displaymath}
     r_{\rm loss} \sim 1 \unit{kpc} \times \sqrt{\frac{1 \unit{GeV}}{E}}
     \sqrt{\frac{K}{0.03 \unit{kpc}^2\unit{Myr}^{-1}}}\;.
\end{displaymath}
The result is an exponential cutoff that depends on the energy,
 i.e. the probability density reads
$d{\cal P}_{\rm rad}/d^3\vec{r}_s\propto 
\exp(-r_s/r_{\rm loss})/r_s$ (see also Sec.4.3 in \citet{Taillet03}).
In the case considered here of sources in
the whole diffusive volume, the normalized probability density is 
given by
\begin{equation}
d{\cal P}_{\rm rad}=
\frac{\exp(-r_s/r_{\rm loss})}{4\pi r_s \,.\,r_{\rm loss}^2} d^3\vec{r}_s\;.
\end{equation}
It is quite different  from the case of a source distribution located in the disk only (see
Eq. (12) in \citet{Taillet03}).
The resulting $f$-volumes (spheres) are given by
\begin{equation}
{\cal P}_{\rm rad}(r<r_{\rm lim}|O)=1-\left(1+\frac{r_{\rm lim}}{r_{\rm loss}}\right)
\exp\left(-\frac{r_{\rm lim}}{r_{\rm loss}}\right)\;.
\end{equation}
It means that sources that contribute to the fraction $f=(50-90-99)$\% of the
detected positrons emitted in the halo are located inside the sphere of radius 
$r_{\rm lim}\approx(1.7-4.8-6.6)\times r_{\rm loss}$.
 For the realistic values of $K(E)$ used above (see also Paper~Ib), we compile in 
Tab.~\ref{tab_positrons} the range covered by $r_{\rm loss}$ at $E=7$~GeV.   
\begin{table}[ht]
\begin{center}
\begin{tabular}{|c||c|c|c|}   \hline
               &$\delta=0.35$	& $\delta=0.60$	& $\delta=0.85$\\\hline\hline
$L=10$~kpc     & 1. 	& .65	 & .48 \\\hline
$L=6$~kpc      & .85 	& .55	 & .41 \\\hline
$L=2$~kpc      & .53 	& .35	 & .26\\\hline
    \end{tabular}
    \caption{The quantity $r_{\rm loss}$ (kpc) is given for three halo sizes $L$ 
    and three diffusion slopes $\delta$ at the total energy $E=7$~GeV.}
  \label{tab_positrons}
\end{center}
\end{table}
Because of the very small scale involved along with the exponential decrease,
$f$-volumes for positrons are expected to be only slightly deformed by 
the dark matter distribution, except for small $\delta$ and large $L$
whose 99\%-volumes extend up to $\sim 7$~kpc.

It is possible now to make a few quantitative comments on the HEAT 
results and on the conclusion of \citet{Baltz02} about this signal. 
They argued that, defining a boost factor related to 
the clumpiness of dark matter, one can accommodate with $e^+$ data without
enhancing too much  the antiproton signal. The point is that antiprotons
come from further than positrons, so that if a clump exists close to us,
its contribution of antiprotons is averaged over a larger zone than 
positrons. A comparison of Figs.~\ref{fig:final1}
and numbers presented above gives a relative distance 
\begin{displaymath}
    r^{\rm origin}_{e^+,~e^-}/
    r^{\rm origin}_{\bar{p},~\bar{d}}\sim 0.1
\end{displaymath}
for all reasonable stationary propagation models.
However, considering large or small $\delta$, the effect of the clumpiness
factor is expected to be different in different propagation models. 
Hence, the enhancement factor for the antiproton signal
used in \citet{Baltz02} should also depend
on the diffusion efficiency, i.e. combination of diffusion
plus convection (that is not considered in the above reference).
To summarize, the relation between SUSY positron and antiproton
signals is not straightforward, if the dark matter halo is clumpy.
Thus it seems a hard task to combine constraints from
these two different signals, unless they are obtained with the same 
analysis. Depending on $\delta$ and $L$, their origin
is more or less local, and the size of the clumps as well as the 
typical distance between the clumps may be of importance.

\section{Summary and conclusions}
This paper analyzes the spatial origin of exotic particles
created from a dark matter profile. We presented the $f$-volumes 
inside which a given fraction of the cosmic rays detected in the Solar 
neighborhood were emitted. 
At high energy ($E \gg 1$ GeV/nuc), the shape of the 
isodensity surfaces  is set by the 
geometry of the diffusive halo, in particular on its height $L$, the 
influence of the side boundary at $r=R$ being small.
We then showed that the $f$-volumes defined are smaller
when spallations and convection are taken into account, but in a very
different way: for particles in the diffusive halo, the wind exponentially
decreases the probability of reaching the Galactic plane, whereas spallations 
have about a null effect on the latter. The parameters $L$ and $2V_c/K$
indicate whether the propagation is convection or escape-dominated.
In Table~\ref{tab5} we summarize the parameters that act as a cut-off 
in various situations.
\begin{table*}[ht]
    \begin{center}
	\begin{tabular}{|c||cc|c|}   \hline
	    Cut-off& Escape-dominated & Convection-dominated &  
	    Losses-dominated\\
	    & ($\chi_{\rm w}\gg1$)	& ($\chi_{\rm w}\ll1$) & 
	    ($e^-e^+\gtrsim$~GeV)\\\hline\hline
	    Halo& $L$		& $L^*\approx 3K/V=3r_{\rm w}/2$ & $
	    \approx 5r_{\rm loss}$\\
	    Radial& $\min(R,3L)$ & $\min(R,3L^*)$ &  $\approx 5
	    r_{\rm loss}$\\\hline
	\end{tabular}
	\caption{Summary of the typical cut-off in $z$ and $r$ directions beyond
	where a cosmic ray cannot originate. The sole parameters that
	determine these cut-offs are $L$ (halo size) and/or 
	$\chi_{\rm w}\equiv r_{\rm w}/L$ -- related to the convective 
	wind $V_c$ {\em via\/} $r_{\rm w}=2K/V_c$, or $r_{\rm loss}$
	related to the effective life-time for positrons and electrons.}
	\label{tab5}
    \end{center}
\end{table*}
Two source distribution for the isothermal dark matter profile
were considered: production related to the density of the source 
(e.g. PBH evaporation), or production related to the square of the density of the sources 
(e.g. SUSY annihilation). The 99\%-volumes are strongly stretched toward the 
Galactic center, corresponding to the maximum of the source distribution. 
This follows from the competition between the
effective source which is maximum at the Galactic
center, and the probability density which steadily decreases
from our position $R_{\odot}$
to reach $\sim 10^{-4}-10^{-5}$~kpc$^{-3}$ for purely diffusive regime
(or even less when convection is included) at the Galactic center.
The fluxes in the Solar neighborhood are found to be far more sensitive to
the dark matter profile in the SUSY case than in the PBH case.
In both cases, the side boundary of the diffusive volume is observed 
to play a negligible role as long as $R\gtrsim 20-30$~kpc.

As a last step, realistic propagation parameters were implemented, 
and the key parameters were found to be the halo size $L$ 
and the diffusion slope $\delta$ (actually $V_c/K_0$).
For the species considered here (antiprotons and antideuterons),
spallations always play a negligible role in the origin. It was found that
this origin is far more local in case of large $\delta$ and small $L$
than in case of small $\delta$ and large $L$. 
Moreover, the shape of the dark matter distribution near the Galactic center 
does not matter so much for the PBH case, whereas it may be crucial for
SUSY annihilating particles.
We emphasized that in any discussion of the annihilation signal in 
charged particles, the propagation parameter $\delta$ or more
precisely, the presence of a Galactic wind, should be 
considered, with the same importance of the parameter $L$ or the 
choice of the dark matter profile.

Two last points are worth noting. First, even though the work presented here 
does not allow a quantitative estimation of the effect of possible
clumpiness of the dark matter halo (for SUSY annihilations), 
we observed that the comparison between the electron and antiproton
SUSY signals should involve a careful inspection of the corresponding
boost factors. Second, whereas the use of B/C-induced propagation parameters
is justified for standard antiprotons (corresponding $f$-surfaces can be 
seen in \citet{Taillet03}), there is no guarantee that these parameters are valid in
the $f$-volumes depicted here.


\section*{Acknowledgments}
This work has benefited from the support of PICS 1076, CNRS and of 
the PNC (Programme National de Cosmologie).

\addtocontents{toc}{\protect\newpage}
\addcontentsline{toc}{part}{Annexes}
\addtocontents{toc}{\protect\vspace{2ex}}
\appendix
\section{Numerical evaluation of the point source solution in Bessel basis}
\label{numerique}
One needs to evaluate numerically point source solutions such as 
\begin{equation}
N^{\rm cyl}_{\delta}(r,z)=\frac{1}{\pi KR} \;
\sum_{i=1}^{n_{\rm tronc}}
\frac{J_0\left(\zeta_i r/R\right)}{\zeta_i J_1^2(\zeta_i)} \frac{\sinh 
\left[\zeta_i(L-z)/R\right]}{\cosh
\left(\zeta_i L/R\right)} 
\label{cyl_pure_diff}
\end{equation}
In the above expression, $(r,z)$ is the position of the $\delta$ source
in polar coordinate and
$R$ is the radial extension of the Galaxy. $N^{\rm cyl}_i(z)$ can be evaluated
for each $i$ and need to be summed till the $n_{\rm tronc}$-th order,
which should formally tend to infinity. For evident reasons, $n_{\rm tronc}$
is chosen to be the smallest possible with the constraint that the rebuilt
series $N^{\rm cyl}_\delta(r,z)$ has reached a good convergence. In the case of
$\delta(\vec{r})$ point source, profiles are singular near the
source and convergence of the series appear to be very slow.
The ansatz depicted in \citet{Taillet03} is useless as soon as
sources are outside the disk. We present below two alternatives
to evaluate this sum.


\subsection{Average value of the oscillating series with $r$}
In analogy with classical
Fourier analysis, resummation of coefficients provide oscillating
behavior around the {\em true\/} value. This can be understood
if we recall that at the $n$-th order, the function
added is $\propto J_0(x\equiv\zeta_n\rho)$: $\rho$ lying in $[0-1]$,
the argument of $J_0$ takes values $x=\{ \zeta_1,\zeta_2,\dots\zeta_n\}$,
i.e. at the $n$-th order, the corrective function has $n$ roots. 
Thus convergence will be more quickly reached if for a given order 
$n_{\rm cutoff}$, instead of evaluating $N^{\rm cyl}_{\delta}(r,z)$, one 
averages
\[
N^{\rm cyl}_{\delta}(r_n,z)=\frac{N^{\rm cyl}_{\delta}(r_{n-1},z)+
N^{\rm cyl}_{\delta}(r_{n+1},z)}{2}\;;
\]
where $r_{n-1}$, $r_n$ and $r_{n+1}$ are ordered realizations of $r$.
The sole condition is that the $\{r_n\}^{n=1,\dots n_{\rm cutoff}}$
belong to the grid $r=\{0,\; R/(2 n_{\rm cutoff}),\; 
2R/(2 n_{\rm cutoff}),\dots, \;R\}$, i.e. $2 n_{\rm cutoff}$ 
linear steps between 0 and 1. To summarize, around the oscillating 
value, if the appropriate step is chosen, it ensures that the averaged 
two points are not both above or below the true value, and furthermore, 
that two opposite extrema of the oscillating function are averaged.


\subsection{Step-like source: $\theta$ function}

An alternative way is to consider solution from a step-like source, 
e.g. $\theta(a-r)$, in order to smooth the problematic behavior 
observed near the origin for the $\delta$ source. With the suitable 
normalization in the source term, i.e.
\[
q_{\theta}(r,z)=\frac{\theta(a-r)}{\pi a^2}\delta(z)\;\;,
\] 
and using the property $\int \rho J_0(\rho)d\rho=\rho J_1(\rho)$, 
it leads to a solution which is equivalent to the delta solution 
$N^{\rm cyl}_{\delta}(r,z)$ -- Eq.~(\ref{cyl_pure_diff}) --, as 
long as the distance $r_o$ of the observer $X_o$ from the source 
satisfies the relation $r_o\gg a$. 
The Bessel coefficients of $\delta$ and $\theta$ solutions are related through
\begin{equation}
N_i^{\theta}(z_o)=2\times\frac{J_1(\zeta_i a/R)}{(\zeta_i a/R)}
\times N_i^{\delta}(z_o)\;.
\label{theta_delta}
\end{equation}
The acceleration of convergence can be understood as, in 
Eq.~(\ref{theta_delta}), the additional term behaves at least
as $1/i$  ($J_1$ is bounded and $\zeta_i\approx i\pi$).
Here $a$ should be taken such as to verify $a/R\ll 1$ (with $R=20$~kpc for 
the Galaxy, one can safely take $a\sim10$~pc).

Thus, a $\theta$-like source slightly underestimates the result close 
to $\vec{r}=\vec{r_s}$, but this zone corresponds to very small volumes that 
add a negligible contribution when one evaluates integrated probabilities.
For practical purposes, both methods (average or $\theta$ source) give the 
desired results with about the same number of Bessel functions, i.e.
$n_{\rm cutoff}\sim 100$.

\bibliographystyle{aa}
\bibliography{ms3378}

\end{document}